# Temperature-Dependent Dynamic Disproportionation in LiNiO$_2$

Andrey D. Poletayev[1,2*], Robert J. Green[3,4], Jack E.N. Swallow[1,2], Lijin An[1,2], Leanne Jones[1,2], Grant Harris[3], Peter Bencok[5], Ronny Sutarto[6], Jonathon P. Cottom[2,7,8], Benjamin J. Morgan[2,7], Robert A. House[1,2], Robert S. Weatherup[1,2*], M. Saiful Islam[1,2,7*]

[1] Dept. of Materials, University of Oxford, Oxford, UK
[2] The Faraday Institution, Harwell Science and Innovation Campus, Didcot, UK
[3] Dept. of Physics and Engineering Physics, University of Saskatchewan, Saskatoon SK, Canada
[4] Stewart Blusson Quantum Matter Institute, Univ. of British Columbia, Vancouver BC, Canada
[5] Diamond Light Source, Harwell Science and Innovation Campus, UK
[6] Canadian Light Source, Saskatoon SK, Canada
[7] Dept. of Chemistry, University of Bath, Bath, UK
[8] Present: Advanced Research Center for Nanolithography, Amsterdam, The Netherlands
* correspondence to andrey.poletayev@gmail.com, robert.weatherup@materials.ox.ac.uk, saiful.islam@materials.ox.ac.uk

## Abstract

Nickelate materials offer diverse functionalities for energy and computing applications. Lithium nickel oxide (LiNiO$_2$) is an archetypal layered nickelate, but the electronic structure of this correlated material is not yet fully understood. Here we investigate the temperature-dependent speciation and spin dynamics of Ni ions in LiNiO$_2$. Our ab initio simulations predict that Ni ions disproportionate into three states, which dynamically interconvert and whose populations vary with temperature. These predictions are verified using x-ray absorption spectroscopy, x-ray magnetic circular dichroism, and resonant inelastic x-ray scattering at the Ni L$_{3,2}$-edge. Charge-transfer multiplet calculations consistent with disproportionation reproduce all experimental features. Together, our experimental and computational results support a model of dynamic disproportionation that explains diverse physical observations of LiNiO$_2$, including magnetometry, thermally activated electronic conduction, diffractometry, core-level spectroscopies, and the stability of ubiquitous antisite defects. This unified understanding of the fundamental material properties of LiNiO$_2$ is important for applications of nickelate materials as battery cathodes, catalysts, and superconductors.

## The unexpected physics of LiNiO$_2$

The broad relevance of nickel-based oxides to applications such as energy storage[1], catalysis[2], and superconductivity[3,4], and the possibility to tune their properties by redox and intercalation[5] motivates a rigorous understanding of the rich underlying physics of these materials[6]. Lithium nickel oxide, LiNiO$_2$, is a widely studied model layered nickelate. In catalysis, LiNiO$_2$ has found use as an effective oxygen evolution catalyst[7]. In Li-ion battery cathodes, the formal Ni$^{3+/4+}$ redox couple offers the highest conventional redox capacity for a given cutoff voltage[1]. Despite this broad interest in LiNiO$_2$, however, to our knowledge, no single model for the electronic structure of LiNiO$_2$ exists that is consistent with all its observed properties.

Since LiNiO$_2$ has previously been comprehensively reviewed in the context of Li-ion batteries[8], here we provide a summary of its key behaviors, including, where relevant, comparisons to other layered alkali metal nickelates $A$NiO$_2$ and rare-earth perovskite nickelates $R$NiO$_3$. The formally 3d$^7$ low-spin ($S$ = ½) configuration of Ni in NiO$_6$ octahedra is orbitally degenerate. Two possible mechanisms for relieving this orbital degeneracy (Figure 1a) are a symmetry-lowering Jahn-Teller distortion or disproportionation[9,10], whereby different Ni ions adopt distinct electronic and geometric local environments. Here we define disproportionation simply as the presence of distinct Ni environments and a process of interconversion between them. Considering other layered nickelates, NaNiO$_2$ exhibits a cooperative and collinear Jahn-Teller distortion[11,12], while AgNiO$_2$ exhibits static disproportionation to multiple distinct nickel environments[13–15]. $R$NiO$_3$ perovskites show similar disproportionation at temperatures below the metal-to-insulator transition, with the oxygens shared unequally between neighboring Ni ions[16–18].

In the case of LiNiO$_2$, both a dynamic non-cooperative Jahn-Teller effect[8,19,20] and a disproportionation of Ni–O bond lengths[21,22] have been proposed, but neither model alone accounts for all the above observations. Here we revisit the mechanism for relieving orbital degeneracy in LiNiO$_2$. We focus on five characteristic behaviors influenced by the local Ni chemistry of LiNiO$_2$:

1. Antisite defects, Ni$_{Li}$, where excess Ni occupies Li sites, are near-impossible to eliminate from LiNiO$_2$, distinguishing it from other layered oxide cathodes[8] and from the sodium analog NaNiO$_2$.

2. LiNiO$_2$ exhibits temperature-activated p-type electronic conductivity[23]. This temperature dependence indicates either Anderson localization or a small-polaron–hopping energy that decreases upon cooling. LiNiO$_2$ with [Ni$_{Li}$] < 3% appears approximately two orders of magnitude more electrically conductive at room temperature than NaNiO$_2$[24], whereas all known polymorphs of AgNiO$_2$ are metallic[25,26].

3. Extended X-ray fine structure (EXAFS) measurements at the Ni K-edge are consistent with distortions of NiO$_6$ octahedra[7,27,28]. These previous studies differ in the direction of the Jahn-Teller distortions assumed when modelling these spectra, and do not consider possible dynamics. Temperature-resolved neutron pair distribution function (PDF) analysis[29] and x-ray diffraction[30] show a gradual transition between cryogenic and room-temperature structures upon heating, rather than an abrupt change of phase.

4. Room-temperature Ni L$_{3,2}$-edge x-ray absorption spectroscopy[22] (XAS) and low-temperature neutron PDF data[29] show substantial differences between LiNiO$_2$ and NaNiO$_2$.

5. The Ni magnetic moments in LiNiO$_2$ are approximately 10% too high for a spin-half 3d$^7$ formal state[31], but regain consistency with a formal Ni$^{3+/4+}$ redox process upon delithiation to 50%, i.e., for Li$_x$NiO$_2$ when $x \leq 0.5$.

Using a combination of ab initio molecular dynamics, three Ni L-edge spectroscopies, and ligand-field multiplet modelling, we show that a dynamic disproportionation model accounts for the five sets of observations above.

Dynamic Disproportionation

We first examine the behavior of Ni environments in LiNiO$_2$ using spin-polarized ab initio molecular dynamics simulations (Methods). At 300 K (Figure 1), the spins of Ni ions are principally distributed across three states: below 0.1 μ$_B$ ($S$ = 0), near 0.86 μ$_B$ ($S$ = ½), or near 1.57 μ$_B$ ($S$ = 1). The spins rapidly convert between these three states via three processes: (i) disproportionation of $S = ½$ Ni ions to $S = 1$ and $S = 0$, e.g., near 420 fs in Figure 1b, (ii) the reverse comproportionation, e.g., near 650 fs, and (iii) exchange, e.g., near 50 fs and 900 fs. All three processes preserve an average formal spin-half state of the Ni ions.

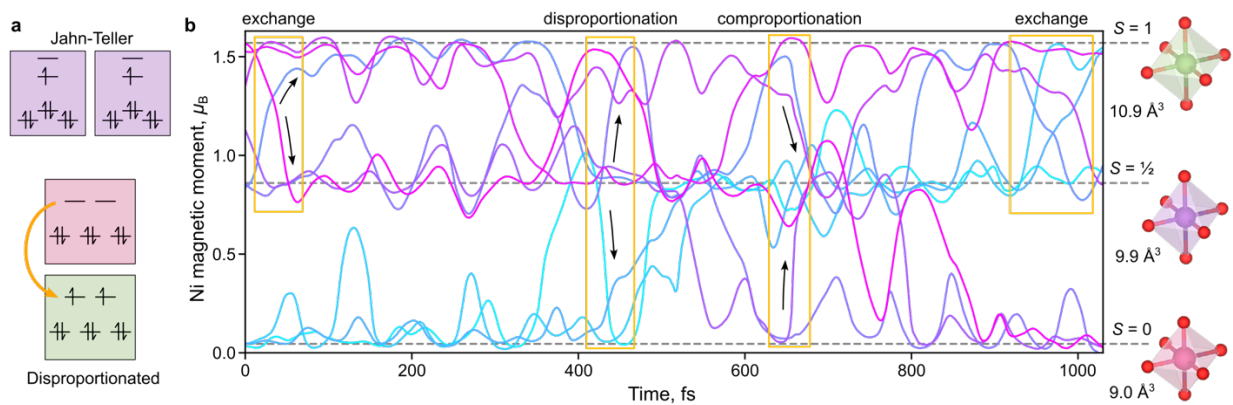

**Figure 1: Ab initio simulation of spin dynamics in LiNiO$_2$.** (a) Simplified schematic of two pathways of relieving orbital degeneracy: Jahn-Teller distortions preserving spin-half electronic structure (purple), and disproportionation (formal electron donation from pink to green). (b) Ab initio molecular dynamics trajectories of Ni spins in a layer containing nine NiO$_6$ octahedra over 1 ps at 300 K, colored by the initial spin (light blue to pink). Exchange, disproportionation, and comproportionation events are highlighted near 50 fs, 420 fs, and 650 fs. NiO$_6$ volumes are annotated for Ni states (green, purple, and pink octahedra).

The limiting case for this three-state system is a structure consisting of three sublattices in the NiO$_2$ layer, each occupied by Ni exclusively in one of the three spin states[21]. In this limiting case, all NiO$_6$ octahedra are somewhat distorted, with the $S = ½$ octahedra showing the strongest Jahn-Teller elongation, as expected. In the three-sublattice structure, all bond distances are below 2.10 Å, consistent with EXAFS[7,27,28]. A small departure from hexagonal lattice symmetry (below 1°) is further consistent with neutron scattering and core-level spectra[21,22,29]. We note the similarity between this limiting structure and the three transition-metal sublattices in Li(NiMnCo)O$_2$[32], noncollinear spin models for hexagonal lattices[33], and the disproportionated structure of AgNiO$_2$[13–15].

We next evaluate the ab initio thermodynamics of spin interconversion and disproportionation in LiNiO$_2$. We construct free energy ($F$) surfaces as $F(s) = - k_B T \ln(p(s))$, where $p(s)$ is the probability distribution of coordinates, $s$, sampled over ab initio trajectories (over 10 ps, Supplementary Information), and $k_B$ and $T$ are Boltzmann's constant and temperature, respectively. As coordinates, $s$, we use Ni magnetic moments and NiO$_6$ volumes, which vary by about 10% with spin states (Figure 1b). The three Ni states appear as basins in the resulting two-dimensional free energy surface (Figure 2a). The magnetic coordinate distinguishes these states more clearly than the NiO$_6$ volume or bond lengths (Supplementary Information), consistent with experiments on perovskite nickelates that demonstrate the primacy of the electronic coordinate[34].

To assess how changes in temperature affect the Ni spin populations, we performed ab initio molecular dynamics at temperatures from 100 K to 600 K. The ab initio free energy surfaces projected onto the spin coordinate (Figure 2b) show that the spin-zero and spin-one states rise in energy from 100 K to 600 K; hence, disproportionation becomes less favorable with heating. Because the local geometric and electronic coordinates are coupled, changes in the relative populations of the three states provide a possible explanation for the experimentally observed gradual evolution of lattice angle with temperature[20,29]. At elevated temperatures, the spin-half state predominates, while the overall rate of transitions between states increases (Supplementary Information).

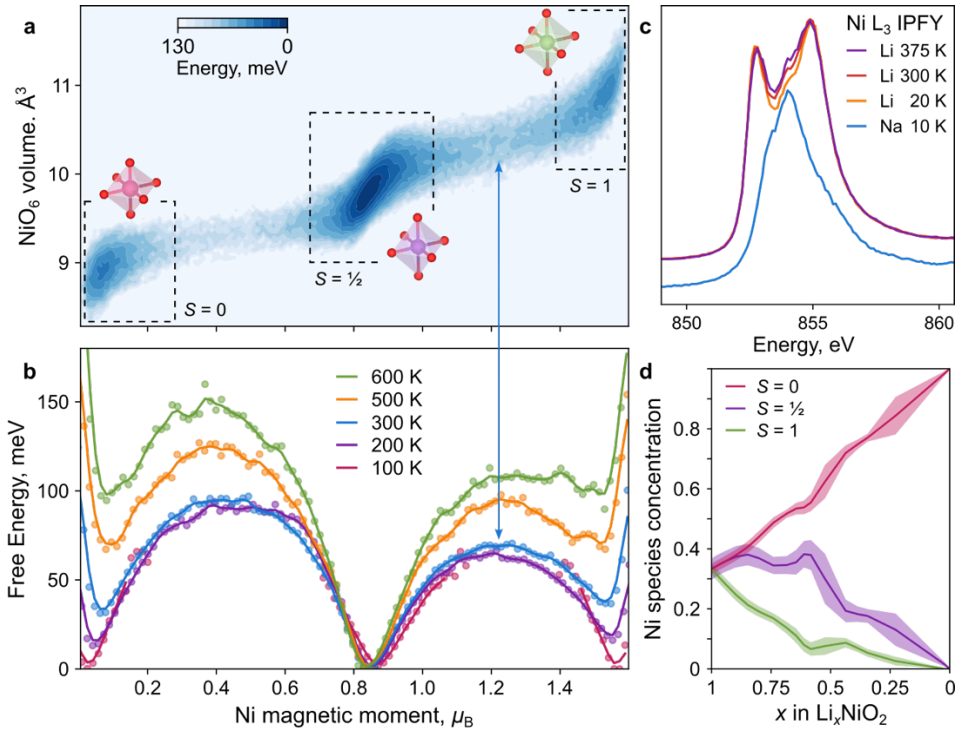

**Figure 2: Temperature dependence of spin disproportionation from simulation and experiment.** (a) Simulated free energy surface at 300 K, versus Ni magnetic moments and NiO$_6$ octahedral volume, with three basins corresponding to spin states highlighted. (b) Simulated free energy profiles versus Ni magnetic moments and temperature. The arrow connecting panels (a) and (b) highlights the saddle point between the $S$ = ½ and $S$ = 1 states. (c) Ni L$_3$ edge x-ray absorption spectra of LiNiO$_2$ in inverse partial fluorescence yield (IPFY) mode[22,35] as a function of temperature. The NaNiO$_2$ spectrum (blue) is offset for clarity. Fits of these spectra to three species are plotted in Extended Data Figure 1. (d) Concentrations of Ni species (green: $S$ = 1, purple: $S$ = ½, pink: $S$ = 0) during delithiation from Monte-Carlo sampling of a DFT-based cluster expansion (Methods). The shaded uncertainty values are ±1 s.e. over eight distinct supercell sizes.

## Spectroscopic Verification

We focus on the qualitative temperature trend for experimental validation. The computational prediction of an increasing fraction of $S$ = ½ Ni species with heating is verifiable if spin states can be distinguished experimentally. Core-level spectra are sensitive to changes in the local electronic states, and we, therefore, measured the temperature evolution of the Ni L$_3$-edge XAS in inverse partial fluorescence yield (IPFY) mode[22,35] (Figure 2c). Two dominant peaks are apparent. Upon heating, these peaks decrease in intensity, while the intensity at the energies between them increases. We therefore discuss these three features in order of increasing energy. First, the low-energy peak is characteristic of NiO-like formally 3d$^8$ species ($S$ = 1). Second, the interpeak energy region that grows in intensity with temperature is at an energy that matches the only peak in the corresponding spectrum of NaNiO$_2$ (Figure 2c, blue). Since NaNiO$_2$ exhibits exclusively a collective Jahn-Teller distortion of S=½ Ni species (Figure 1a), we ascribe this middle energy region to $S$ = ½ Ni species in LiNiO$_2$. Third, the high-energy peak could plausibly arise from a lower-spin state such as $S$ = 0.

This evolution of the Ni L-edge is analogous to that observed in rare-earth perovskite nickelates RNiO$_3$, where double-peaked edge shapes morph into a broad and flat edge with heating across the metal-to-insulator transition [17,18]. The overall temperature evolution of LiNiO$_2$ XAS spectra is weaker than predicted by the increase in the relative proportion of $S = ½$ Ni species with temperature in our ab initio simulations (Figure 2b), but the two are qualitatively consistent. We conclude that LiNiO$_2$ exhibits Ni-disproportionation that is both dynamic and temperature-dependent. Notably, if a Jahn-Teller distortion, collective or not, exclusively accounted for the low-temperature local geometry of LiNiO$_2$, or if disproportionation were only activated with heating, then a stronger semblance to the NaNiO$_2$ spectra would be expected at low temperature, and the evolution of the spectra should be reversed, i.e., the low- and high-energy peaks would be expected to grow with heating.

Having experimentally validated our model of three-fold dynamic disproportionation, we use this model to predict Ni speciation upon delithiation, as occurs during battery cycling. Using grand canonical Monte-Carlo simulations (Figure 2d), we predict that during the first half of delithiation (Li content $x > 0.5$ in Li$_x$NiO$_2$), the high-spin Ni species are first to be oxidised, corresponding to net formal Ni$^{2+/4+}$ redox. For $x < 0.5$, the more expected Ni$^{3+/4+}$ redox dominates, as reported from bulk-sensitive x-ray Raman scattering[36]. This predicted sequence of redox events is also consistent with magnetometry[31].

## Spectral shapes of Ni species

To understand the origin of the observed changes in spectral features, we perform ligand-field charge-transfer multiplet simulations[37]. Accounting for unequal Ni–O bond lengths arising from both the NiO$_6$ volume differences and the Jahn-Teller distortion predicted from simulation (Methods) affords a first-principles prediction of state-specific spectral shapes for LiNiO$_2$ and NaNiO$_2$ (Figure 3 and Extended Data Figure 1). Our predicted spectra reproduce the experimentally observed TEY and IPFY spectra for both materials. In LiNiO$_2$, the $S = 1$ and $S = 0$ components account for the low- and high-energy L$_3$ edge peaks, respectively. This picture is consistent with both a partial disproportionation and with the usual small Ni excess in LiNiO$_2$, which contributes to the $S = 1$ feature (3–5% in IPFY; Figure 3b and Extended Data Figure 1). For the spectra in Figure 2c, the proportion of $S = ½$ species grows from 35% at 20 K to 41% at 375 K (Extended Data Figure 1). Even though a precise quantitative agreement may be beyond the accuracy of the predictions of density-functional theory (DFT), our experimental results are consistent with disproportionation in LiNiO$_2$ and confirm the increase in the proportion of $S = ½$ ions with temperature. We discuss the sensitivity of computational predictions further in the Supplementary Information.

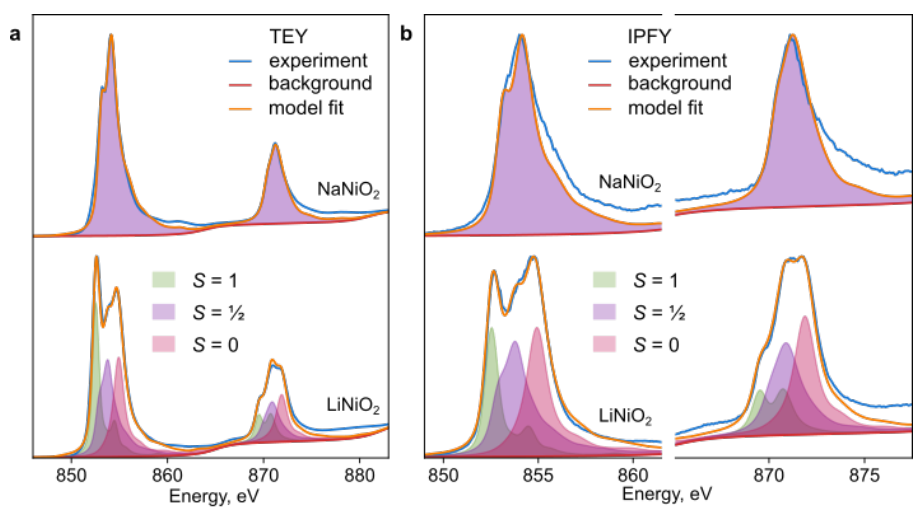

**Figure 3: Decomposition of Ni L edge spectra of NaNiO$_2$ and LiNiO$_2$.** (a) TEY spectra, (b) IPFY spectra. NaNiO$_2$ (top) was measured at 10 K modelled exclusively using the spin-half component (Methods). LiNiO$_2$ (bottom) was measured at 6 K and fit to 42%-35%-23% $S = 1$, $S = ½$, and $S = 0$ components, respectively, (TEY) or 33%-39%-28% of the same (IPFY). The IPFY L$_2$ edge was rescaled due to saturation.

The calculated partial densities of states for the three Ni species (Extended Data Figure 2b) verify that both $S = 1$ and $S = ½$, but not $S = 0$, species contribute to the valence band edge. As with other high-valence Ni compounds[38], strong covalency is predicted here for the $S = ½$ (mostly d$^8$L̲) and $S = 0$ (mostly d$^7$L̲ and d$^8$L̲$^2$) species (Extended Data Figure 2a). A key novelty of our work is the confirmation that these formally high-valence species are present in the pristine, fully lithiated material. Therefore, we next verify the detection of $S = 1$ and $S = 0$ species with x-ray magnetic circular dichroism (XMCD) and resonant inelastic x-ray scattering (RIXS), respectively.

XMCD was performed at the Ni L$_{3,2}$-edge under 8 T applied field (Figure 4). Circular dichroism is specifically sensitive to unpaired electrons at the Ni centers and can elucidate the competing degrees of charge transfer and covalency[39]. The XMCD spectra thereby assist in constraining the charge transfer multiplet calculations[40]. The L$_3$ XMCD spectra differ between LiNiO$_2$ and NaNiO$_2$ (Figure 4), mirroring the different x-ray absorption spectra, above. The LiNiO$_2$ L$_3$ XMCD spectrum has a maximum at about 1 eV lower energy and exhibits a sign change near 855 eV in IPFY. The disproportionation model reproduces the XMCD spectra of both compounds in TEY and IPFY modes. The broader dichroism features of NaNiO$_2$ versus the $S = ½$ Ni species in LiNiO$_2$ are consistent with NaNiO$_2$ exhibiting a stronger Jahn-Teller distortion; XMCD (Figure 4) appears more sensitive to Jahn-Teller distortions than x-ray absorption (Figure 3), where the $S = ½$ shapes are more similar for the two materials. The computed signature of $S = 1$ Ni species in LiNiO$_2$ (green in Figure 4) includes a sign change characteristic of spinel Ni$^{2+}$, as seen for NiFe$_2$O$_4$ spinel[41,42]. This feature accounts for the lower-energy L$_3$ peak and sign change of the dichroism in LiNiO$_2$ relative to NaNiO$_2$, especially in the more bulk-sensitive IPFY mode. The presence of about 10% excess reduced Ni species near the surface of LiNiO$_2$ observed in TEY mode relative to IPFY (Figure 3) prevents a more quantitative assignment of the LiNiO$_2$ TEY XMCD spectrum. Nevertheless, the differences between XMCD spectra of the two materials are consistent with the presence of $S = 1$ Ni in bulk LiNiO$_2$ due to disproportionation.

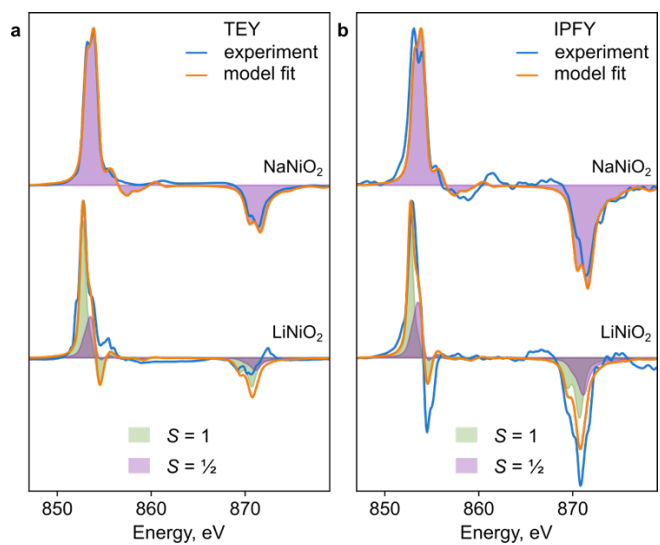

**Figure 4: Ni L edge X-ray magnetic circular dichroism (XMCD) of LiNiO$_2$ and NaNiO$_2$.** (a) TEY, (b) IPFY. NaNiO$_2$ (top) was measured at 10 K and 8 T field, LiNiO$_2$ (bottom) was measured at 10 K and 8 T field. The models (orange) are fit using the same compositions as in Fig. 3. The XMCD signature of the spin-zero component is negligible. The calculated L$_2$ IPFY XMCD was scaled up by the same factor as the linear L$_2$ spectra in Fig. 3b. Raw spectra: Figure S1.

The $S = 0$ species does not possess an XMCD signature. We, therefore, verify its presence using the added dimension of inelastic energy loss in RIXS. The L$_3$ RIXS map of LiNiO$_2$ (Figure 5a) includes two features that distinguish it from prior reports of nickelate RIXS[18]. First, there is intensity approaching the elastic line near 853.5 eV, which is about 1 eV higher than in metallic NdNiO$_3$[18]. Second, features near 2 eV and 6 eV loss at 854 eV–855 eV have not, to our knowledge, previously been reported. The fluorescence feature (dotted diagonal in Figure 5a) extends to < 1 eV loss at 852.0 eV, suggesting that $LiNiO_2$ possesses a nonzero optical bandgap[18]. To interpret the RIXS maps, we extended the same Anderson impurity model of three-fold disproportionation as used to interpret the XAS and XMCD data, without any additional optimization (Supplementary Information), and computed RIXS maps for three Ni species, weighted as for IPFY (Figure 3b). We discuss loss spectra at three incident photon energies, denoted (i)-(iii) in Figure 5a.

At 852.5 eV (Figure 5b, (i)), the main contributions come from $S = 1$ Ni species. This energy loss spectrum is similar to spectra of materials containing $d^8$ states, such as the binary oxide $NiO^{43}$ and perovskite $NdNiO_3^{18}$. Here, the model slightly over-estimates the crystal field splitting and reproduces the experimental spectrum with a slight shift to higher loss energies. However, surface reduction and the overlap of the main $d$–$d$ excitation with fluorescence near 1 eV loss may contribute to the mismatch here.

At 853.5 eV (Figure 5b (ii)), low-loss excitations attributed to $S = 1$ and $S = ½$ Ni species extend to the elastic line, consistent with the presence of states just below and above the Fermi level attributable to both species in DFT calculations (Extended Data Figure 2b). Broad states above 4 eV loss, above the fluorescence feature (seen at 2 eV loss for this photon energy) and attributable to $S = ½$ Ni species, likely arise from charge-transfer excitations, consistent with the $d^8\underline{L}$ contribution to its ground state.

Finally, at 854.5 eV (Figure 5b, (iii)) our model attributes the feature at 6 eV loss exclusively to $S = 0$ Ni species. The high energy loss of this component suggests it is also of charge-transfer origin, but this feature is not present in $NdNiO_3^{18}$. Strong contributions of the $S = 0$ species are also evident at 2 eV loss. As at lower photon energies, the model slightly overestimates energy loss, but reproduces the major spectral features. We conclude that RIXS specifically detects the presence of $S = 0$ Ni species and confirms disproportionation in LiNiO$_2$. Additional weak transitions at 1-2 eV loss above 856 eV (Figure 5a) are also attributable to the $S = 0$ Ni species (Supplementary Information). While additional fine-tuning of the charge-transfer multiplet model parameters is possible based on the RIXS spectra, we forego this here because of the contributions of reduced surface layers, which likely resemble NiO, to the main $S = 1$ feature, as in TEY (Figure 3a).

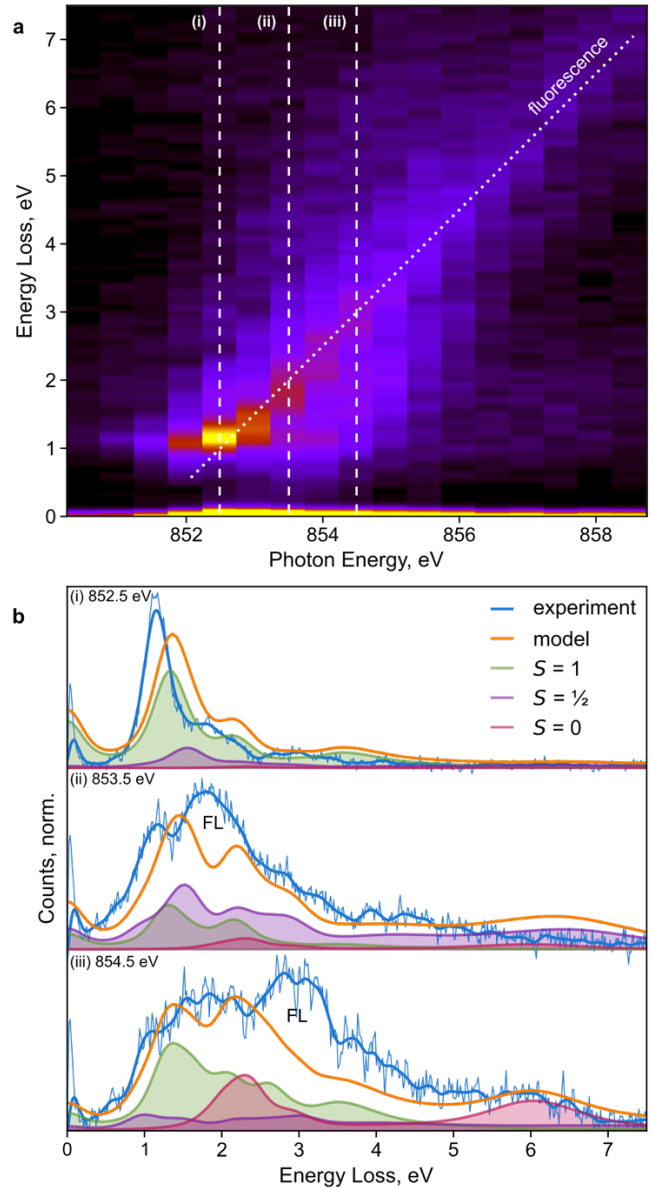

**Figure 5: Ni L edge resonant inelastic x-ray scattering (RIXS) of $LiNiO_2$.** (a) RIXS intensity map measured across the $L_3$ edge at 20 K, (b) energy loss spectra (blue) at incident photon energies (i) 852.5 eV, (ii) 853.5 eV, and (iv) 855.0 eV compared to calculated loss spectra (orange). Calculated spectra were normalized to 85% of the maximum experimental intensity to account for the fluorescence feature (FL). Relative compositions of nickel species (green, purple, pink) were the same as for IPFY (Figure 3b). Full calculated d-d and charge-transfer intensity maps are shown in Figure S5.

## Consistency with observables

The model of dynamic and temperature-dependent disproportionation presented here is consistent with the five observed behaviors of LiNiO$_2$ detailed above, summarized in order:

1. Additional S=1 Ni ions, arising from disproportionation with the NiO$_2$ layers, are predicted to stabilize Ni$_{Li}$ defects through favorable antiferromagnetic (AFM) interactions (Supplementary Information), explaining the ubiquity of this antisite defect.
2. Activated electronic conduction plausibly arises from exchange (Figure 1a) between $S = ½$ and $S = 1$ Ni species. Indeed, the simulated free energy at the saddle point (Figure 2b) is close to half of the activation energy of electronic conductivity[23]. The increase in this saddle-point energy as the $S = ½$ state predominates at high temperatures (Figure 2b) is consistent with increased activation of conductivity upon heating. Electron and hole polarons can be localized in the disproportionated structures, supporting a correspondence between formal spin and charge states (Supplementary Information). In contrast, in NaNiO$_2$, the collective Jahn-Teller distortion precludes the exchange of spin states and reduces electronic conductivity.
3. The small distortion of the LiNiO$_2$ unit cell is consistent with that of the limiting three-fold disproportionated cell, while the gradual decrease in this unit cell distortion with heating[20,29] is consistent with the gradually increasing proportion of $S = ½$ species.
4. Previously not reported XMCD (Figure 4), Ni L$_3$ edge RIXS (Figure 5), and temperature-resolved XAS (Figure 2c) data provide strong experimental evidence for the disproportionation of Ni species in LiNiO$_2$. Accounting for $S = 0$ and $S = 1$ species affords an interpretation consistent across the Ni L-edge spectroscopies of LiNiO$_2$ (Figures 3-5) and of the spectroscopic differences between LiNiO$_2$ and NaNiO$_2$. These observations, combined with charge-transfer multiplet modelling, confirm a negative charge-transfer regime for both compounds[44], but highlight their distinct mechanisms of relieving degeneracy (Figure 1a).
5. The presence of $S = 1$ Ni species in bulk LiNiO$_2$ until 50% delithiation accounts for the increased Ni magnetic moments relative to those expected from formal Ni$^{3+}$ in LiNiO$_2$[31] and is further consistent with bulk-sensitive x-ray Raman scattering[36].

## Conclusions

We have identified temperature-dependent dynamic disproportionation as the mechanism relieving orbital degeneracy in the archetypal layered nickelate LiNiO$_2$. Our results support a unified model where Ni species in LiNiO$_2$ exhibit three states with formal spins $S = 0$, $S = ½$, and $S = 1$ (which correspond to the formal oxidation states Ni$^{4+}$, Ni$^{3+}$, and Ni$^{2+}$, respectively) and interconvert between these on a picosecond timescale. We have verified this behavior with characterization of the nickel L-edge using XAS, XMCD, and RIXS. The low-spin species exhibit strong Ni-O covalency within a charge-transfer multiplet model. These results enable the fingerprinting of the nickel L$_{3,2}$ absorption edges based on first-principles calculations. The temperature dependence and dynamic nature of the disproportionation extend earlier models[21,22] and allow for consistency with a diverse set of experimental observables:

thermally activated electronic conductivity[23], local structure from neutron diffraction[29], magnetometry[31], stabilization of antisite Ni excess defects, and, more generally, the gradual changes in the properties of LiNiO$_2$ with heating and delithation. Overall, our unified picture of Ni behavior will advance characterisation and understanding of the physics of nickelate materials for a range of applications, including rechargeable batteries, catalysis, and superconductivity.

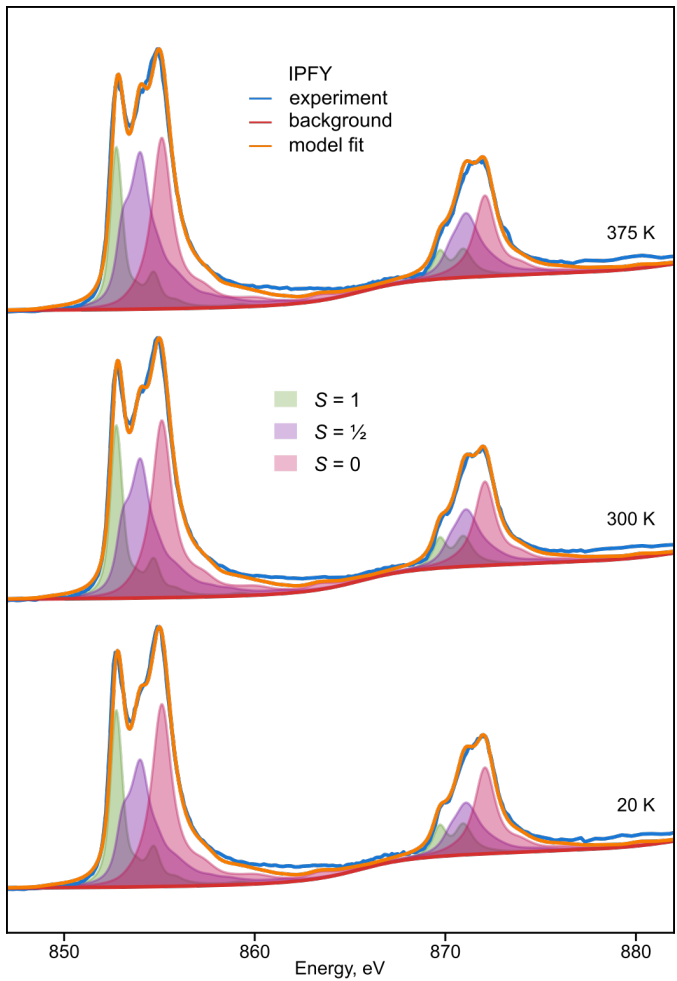

**Extended Data Figure 1: Temperature dependence of the IPFY spectra of LiNiO$_2$.** The fitting of the IPFY spectra in Figure 2c with $S$ = ½ fractions 35%, 37%, and 41% at 20 K, 300 K, and 375 K, respectively, and the excess of $S$ = 1 species over the $S$ = 0 species is 3% at all temperatures. The difference between the fits to these spectra, collected at CLS, and those in Figure 3, collected at DLS, is 2% in both parameters at 300 K. The increasing smoothness of the LiNiO$_2$ spectra at the energies where the spectrum of the $S$ = ½ species has a maximum, in the middle of the edges, towards elevated temperature is further consistent with dynamic interconversion.

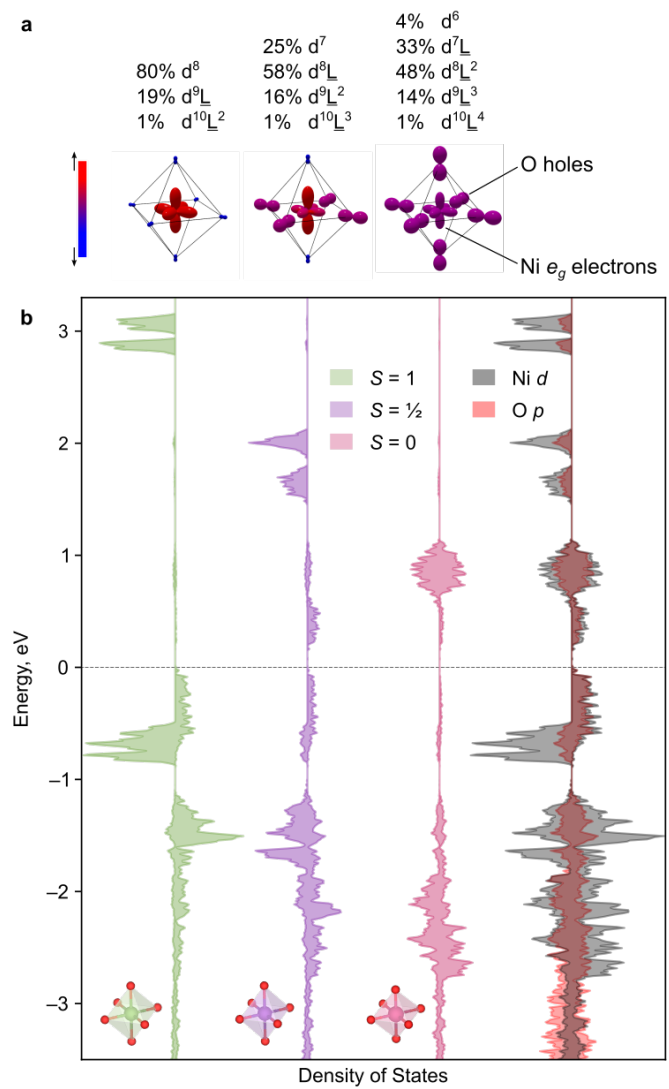

**Extended Data Figure 2: Models of three Ni species in LiNiO$_2$.** (a) Bond length and charge-transfer multiplet components for each species, and Ni $e_g$ electron and ligand hole density matrix plots. Salient features are high ionicity and low ligand hole content for the ($S = 1$), preferred occupancy of the $3z^2$–$r^2$ orbitals and bonding anisotropy consistent with Jahn-Teller distortions ($S = ½$), and spin depolarization and ligand hole contributions ($S = 0$). (b) Projected densities of Ni 3d states for $S = 1$ (green), $S = ½$ (purple), and $S = 0$ (pink) Ni species near the Fermi level. Right: total Ni 3d (grey) and O 2p (red) densities of states. The spin depolarization of the $S = 0$ states near 1 eV is consistent with the density matrix from ligand field calculations.

## Methods

*Sample preparation*

Uncoated, polycrystalline LiNiO$_2$ powder was obtained from BASF. NaNiO$_2$ was prepared in house by a solid-state reaction. Appropriate molar amounts of Na$_2$CO$_3$ and NiO were ground together in a pestle and mortar, pressed into a pellet, and then heated at 650 °C under flowing O$_2$ for 12 hours. The heating and cooling rates were controlled at 10 °C min$^{-1}$. Powder X-ray diffraction data were collected for LiNiO$_2$ and NaNiO$_2$ on Cu-source Rigaku diffractometers. GSAS-II software was used to perform the Rietveld refinement analysis.

*IPFY XAS and XMCD*

Temperature dependent XAS measurements of LiNiO$_2$ were performed at the REIXS beamline of the Canadian Light Source (CLS). Samples were transported to the facility in sealed vials under argon atmosphere, pressed onto carbon tape on copper sample plates under argon atmosphere in a glovebox, and loaded into the x-ray experimental chamber without exposure to atmosphere. Measurements were performed at 20-375 K at pressures below 10$^{-9}$ mBar. The incident beam was horizontally polarized and the normal of the sample plate was aligned with the beam. XAS was collected with TEY by monitoring sample drain current, and IPFY and PFY using a silicon drift detector with ~70 eV resolution. The silicon drift detector was positioned at an angle approximately 60 degrees from the sample normal.

Temperature dependent XMCD and XMLD measurements of LiNiO$_2$ and NaNiO$_2$ were performed in IPFY mode, with simultaneous TEY and FY detection at both the O K and Ni L$_{3,2}$ edges on the high-field magnet end station at the I10 beamline, Diamond Light Source, UK. Powder and electrode samples were mounted onto a copper sample plate using carbon tape in an inert Ar-filled glovebox atmosphere, before being transported directly to the chamber in an Ar-filled sealed transfer vessel (avoiding exposure to air). Measurements were performed at 6-300K under ultra-high vacuum conditions. The incident beam was directed at a 60° angle to the normal of the sample plate. FY was acquired in the same 60° back-scattering geometry using a Si diode with an Al cover to filter out emitted electrons, mounted in front of the beam entrance port. IPFY was recorded with a four-element Vortex Si drift detector mounted at 90° to the incoming beam (30° to sample normal). XMCD and XMLD measurements were performed at 8 T and collected through the individual detection of right ($\sigma_r$) and left ($\sigma_l$) circular polarizations, or linear horizontal ($\sigma_h$) and vertical ($\sigma_v$) polarizations. The powdered form of the samples means we expect measured signals to be anisotopically averaged, i.e., significant orientation effects are not expected, although this likely reduces the observed extent of dichroism. Both O K-edges and Ni L$_{3,2}$-edges were measured in the continuous scanning mode of the monochromator, with an energy step size of 0.1 eV. All data was divided by the I$_0$ signal to remove top-up intensity spikes and energy-dependent intensity variations associated with the beamline. IPFY data was processed by summing the O emission signal over the incident energy range and following the procedure of Achkar et al.[35] The pre-edge average background was subtracted, and remaining intensity normalized by the post-edge average.

*Ni L$_3$ edge RIXS*

Ni L$_3$-edge RIXS spectra were measured at a temperature of 20 K at the I21 beamline, Diamond Light Source[45]. The incident energy range was 849-859 eV in 0.5 eV steps. Samples were transferred to the spectrometer using a vacuum-transfer suitcase to avoid air exposure and were pumped down to ultra-high vacuum (UHV) and left to fully degas overnight.

*Computational: density-functional theory, ab initio molecular dynamics, cluster expansion*

DFT simulations were carried out using the projector-augmented wave method[46–48] in the VASP package[49,50] using the meta-GGA functionals SCAN[51] and r$^2$SCAN[52] and forgoing empirical parameters such as a Hubbard *U* correction or the fraction of exact exchange. The revised Vydrov-van Voorhis (rVV10) non-local dispersion correction was applied. As we were not aware of the accurate parameterization of the rVV10 correction for r$^2$SCAN[53] until substantially after running extended ab initio molecular dynamics simulations using the parameterization for SCAN (b=15.7, c=0.0063)[54], and the favorability of disproportionation was sensitive to the functional over the dispersion correction, the molecular dynamics were not re-run. Static calculations were completed with the parameterization for r$^2$SCAN (b=11.95, c=0.0063), 700 eV plane-wave cutoff, and 0.25 Å$^{-1}$ *k*-point spacing. Energies and forces were relaxed to 10$^{-5}$ eV and 10$^{-2}$ eV/Å, respectively, or better. Ab initio molecular dynamics (AIMD) simulations used a Γ-centered 2×2×2 *k*-point mesh, 2 fs time steps, constant-volume (NVT) ensemble, Nosé-Hoover thermostat with a time constant of 40 steps, electronic convergence of 10$^{-4}$ eV, and the preconditioned conjugate gradient algorithm (VASP ALGO=A), unless specified otherwise.

To identify the states of the Ni we use local spin densities, *S*, as calculated in VASP. This descriptor gives a relatively unambiguous assignment for each Ni without estimating formal charges from the full electronic density in post-processing. The first picosecond of every AIMD run was excluded from analyses for thermostat equilibration. The simulations at 100 K and 200 K, where sampling transitions between Ni states required long trajectories, were initialized by cooling from 300 K over 500 fs or longer. AIMD simulations with a Ni$_{Li}$ defect were initialized with the starting spin of the antisite Ni set to -2 $\mu_B$, and all others as default (1 $\mu_B$). The trajectories of the nickel spins were binned into *S* = 0, *S* = ½, and *S* = 1 states by milestoning[55] with cutoffs of 0.2 $\mu_B$, 0.7 $\mu_B$, 1.02 $\mu_B$, and 1.4 $\mu_B$. A control simulation in the isobaric (NPT) ensemble was carried out with the Langevin thermostat coupled only to the Li atoms at 12 ps$^{-1}$ to avoid perturbing the dynamics of Ni-O bonding.

A decorated cluster expansion of defect-free LiNiO$_2$ was trained to predict the nickel speciation on delithiation[56]. Reference structures for training were chosen to be large enough to allow for disproportionation should that be favorable, and pre-distorted for accelerating relaxation. The DFT settings for reference structures were as for static calculations above, although some relaxations were shortened when clearly approaching convergence due to the reduced requirements on precision for the purposes of the cluster expansion. The root mean squared errors (RMSE) were 4.6 meV/f.u. over the training set and 5.6 meV/f.u. over the hold-out validation set. Charge-neutral grand canonical Monte-Carlo (CNGCMC) sampling was used to estimate the nickel speciation at all states of delithiation (Figure 2d), with spin states used as formal charge states for nickel. To mitigate the effects of commensurate lattice orderings[57] on predicted speciation, eight different supercell sizes were averaged. For each chemical potential of Li vacancies, the CNGCMC runs were initialized at 1000 K, cooled to 100 K for finding the ground state, heated to 500 K, and sampled for 10$^6$ steps, with the first half of those discarded. The concentrations of Ni species were averaged over supercells for each chemical potential of Li vacancies[58]; chemical potentials of Ni species were kept at zero relative to each other. A more detailed study of delithiation in LiNiO$_2$ and the limitations of the cluster expansion formalism is the subject of follow-on work.

Defect formation energies were calculated only for charge-neutral structures from relaxed defect-free and defect-incorporating cells[59–62]. The chemical potentials of the elements at synthesis conditions were calculated from the energies of the reference phases[62–64]. At the typical conditions of synthesis—1 atm O$_2$ pressure and 700 °C—the chemical potential of oxygen is $\mu_O$ = -1.065 eV, which determines $\mu_{Li}$ = -2.962 eV and $\mu_{Ni}$ = -1.379 eV. We account for the antiferromagnetic–paramagnetic transition of NiO at its Néel temperature by taking the energy of paramagnetic NiO as the average of computed AFM and FM energies.

*Multiplet ligand field theory modelling of the Ni L$_{3,2}$ edge spectroscopies*

The nickel L$_{3,2}$-edge multiplet ligand field theory (MLFT) simulations were performed using the many-body code Quanty.[65] The simulation was implemented using a single-cluster NiO$_6$ Hamiltonian of $O_h$ symmetry for $S$ = 0,1 and $D_{4h}$ symmetry for $S$ = ½. The Ni 2p, Ni 3d, and ligand shells are explicitly included. For all calculations, Slater integrals are scaled to 80% and 85% for the initial and final Hamiltonians, respectively. Additionally, onsite ligand energy shifts of $T_{pp}$=±0.75 eV were applied to the ligand orbitals of $e_g$ (+) and $t_{2g}$ (-) symmetry.

A charge transfer energy of Δ = -0.5 eV assumed for the 3d$^7$ $S$ = ½ Ni, as used by Green et al.[37] This charge transfer energy, along with a Coulomb interaction energy of $U_{dd}$ = 6 eV, leads to charge transfer energies of 5.5 eV and -6.5 eV for the $S$ = 1 (3d$^8$) and $S$ = 0 (3d$^6$) clusters, respectively. A core-valence Coulomb interaction parameter of $U_{pd}$ = 7 eV was used, which is the standard ~1 eV larger than $U_{dd}$. Hopping integrals and crystal field energies are obtained directly from bond lengths using Harrison's formulas[37,66], and hopping integrals were scaled by 80% in the XAS final state[37]. The DFT-determined bond lengths were used for the three sites in LiNiO$_2$. For NaNiO$_2$, bond lengths of 1.93 Å and 2.16 Å for x/y and z bonds were used, respectively[67–69], which yields a slightly larger Jahn-Teller distortion than for the LiNiO$_2$ $S$ = ½ site geometry. The charge transfer energies, hopping integrals, and crystal field potential energies are listed below for all calculations.

$S$ = 1 calculation (eV): Δ= 5.5, crystal field 10D$_q$= 0.71, hopping integrals V$_{eg}$ = 2.63, V$_{t2g}$ = 1.52.
$S$ = ½ calculation (eV): Δ = -0.5, 10D$_q$ = 0.78 with Jahn-Teller splitting of Δ$_{eg}$= 0.15 and Δ$_{t2g}$= 0.10. Here, Δ$_{eg}$ denotes the difference between the $x^2 – y^2$ and $3z^2 – r^2$ onsite energies, and Δ$_{t2g}$ the difference between the $xy$ and $xz/yz$ onsite energies (eV): V$_{3z2-r2}$ = 2.43, V$_{x2-y2}$ = 3.33, V$_{xz/yz}$ = 1.41, V$_{xy}$ = 1.93.
$S$ = 0 calculation (eV): Δ = -6.5, 10D$_q$ = 0.93, V$_{eg}$= 3.456, V$_{t2g}$ = 2.004.
$S$ = ½ calculation for NaNiO$_2$ (eV): Δ = -0.5, 10D$_q$ = 0.70 with Jahn-Teller splitting of Δ$_{eg}$ = 0.19 and Δ$_{t2g}$ = 0.12. V$_{3z2-r2}$ = 2.02, V$_{x2-y2}$ = 3.17, V$_{xz/yz}$ = 1.17, V$_{xy}$ = 1.84.

Acknowledgements

The authors acknowledge funding from the UK Faraday Institution (faraday.ac.uk; EP/S003053/1, FIRG016, FIRG024) and the European Research Council (ERC) under the European Union's Horizon 2020 research and innovation programme (EXISTAR, grant agreement No. 950598). B.J.M. acknowledges support from the Royal Society (URF/R/191006). R.S.W. acknowledges a CAMS-UK Fellowship through the Analytical Chemistry Trust Fund and a UKRI Future Leaders Fellowship (MR/V024558/1). R.J.G. and G.H. acknowledge funding from the Natural Sciences and Engineering Research Council of Canada (NSERC). The authors acknowledge the HEC Materials Chemistry Consortium (EP/R029431) for the use of Archer2 high-performance computing (HPC) facilities. The authors also acknowledge the Faraday Institution's Michael HPC resource. We acknowledge Diamond Light Source for time on beamlines I10 and I21 under proposals MM33062 and MM30644-1, and Dr. Stefano Agrestini, Dr. Mirian Garcia-Fernandez, and Dr. Ke-Jin Zhou for assistance with the RIXS measurements. We acknowledge the support of the Royal Academy of Engineering under the Research Fellowship scheme. Part of the research described in this paper was performed at the Canadian Light Source, a national research facility of the University of Saskatchewan, which is supported by the Canada Foundation for Innovation (CFI), the Natural Sciences and Engineering Research Council (NSERC), the Canadian Institutes of Health Research (CIHR), the Government of Saskatchewan, and the University of Saskatchewan. A.D.P. is grateful to Dr. Pezhman Zarabadi-Poor and Dr. Gregory Rees for insightful discussions.## Author Contributions

Initial investigations of $LiNiO_2$ were carried by J.P.C. and extended to AIMD by A.D.P. with advice and supervision from M.S.I. and B.J.M. The temperature-dependent XAS experiments were proposed by A.D.P. and carried out by R.J.G. and R.S. (CLS). The XMCD experiments were proposed by R.J.G., J.E.N.S., R.S.W., and A.D.P., and carried out by A.D.P., J.E.N.S., L.A., and L.J. with P.B. (DLS). The RIXS experiments were proposed by A.D.P. and R.A.H. and carried out by R.A.H. with S.A., M.G.-F., K.-J.Z. Samples were prepared by R.A.H., L.A., J.E.N.S., and L.J. The charge-transfer multiplet modelling was carried out by G.H. and R.J.G. A.D.P. led the writing of the manuscript with input and contributions from all authors.

# Supplementary Information

## Characterization and experimental controls

*TEY, TFY, IPFY comparison of LiNiO$_2$ and NaNiO$_2$*

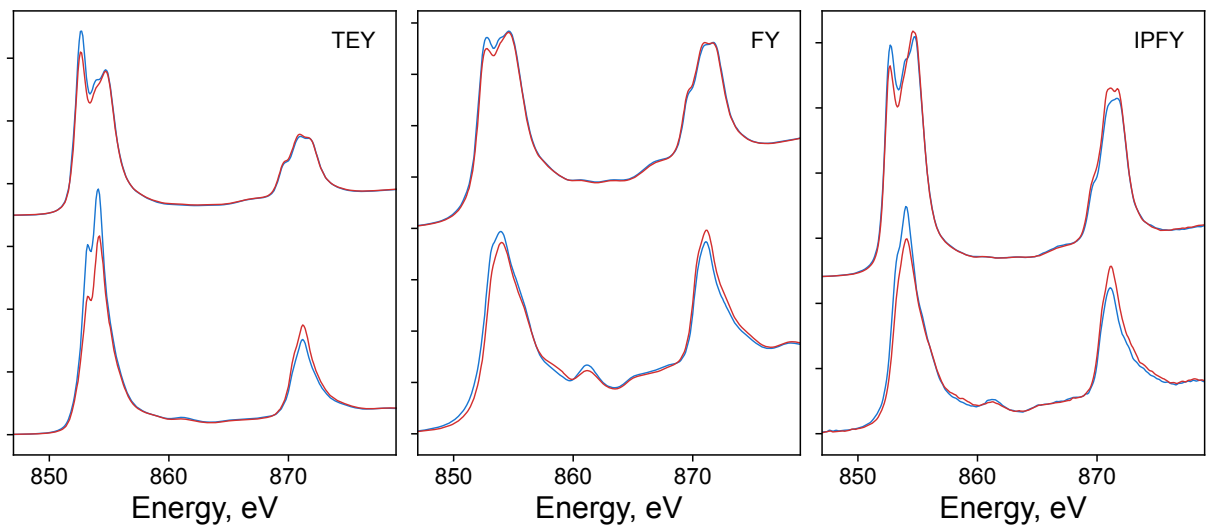

**Figure S1: Ni soft x-ray absorption spectroscopy using three detection modes.** (a) total electron yield, (b) fluorescence yield, (c) inverse partial fluorescence yield via emission at the O K-edge. Blue: σ$_r$ polarisation, red: σ$_l$ polarisation. Top: LiNiO$_2$, 6 K, 8 T. Bottom: NaNiO$_2$, 10 K, 8 T.

*Absence of XMLD at the Ni edge and absence of dichroism at the O K edge*

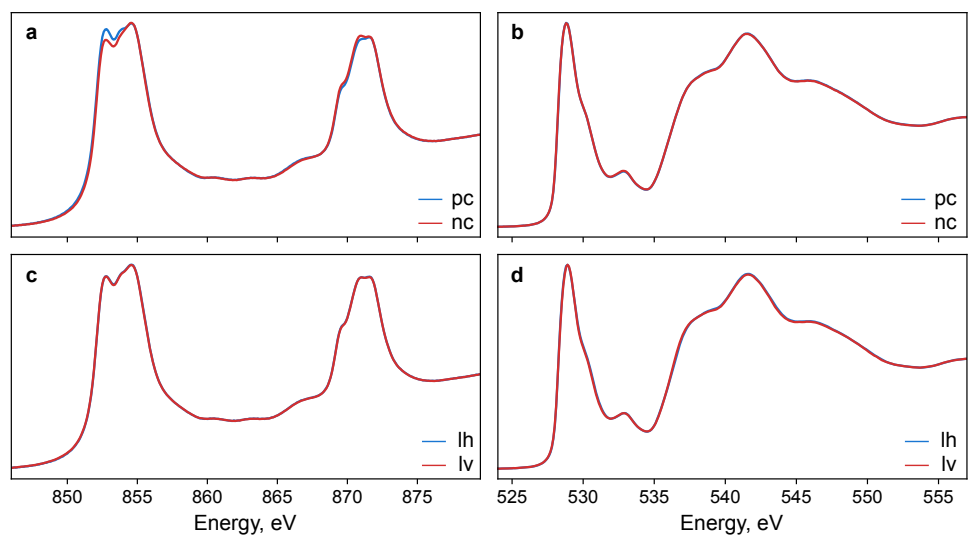

**Figure S2: x-ray circular and linear dichroism in LiNiO$_2$.** (a) circular, Ni L-edge, (b) circular, O K-edge, (c) linear, Ni L-edge, (d) linear, O K-edge. All: fluorescence yield, 6 K, 8 T. Only circular dichroism is observable, and only at the Ni L-edge.

*XRD of LiNiO₂ and NaNiO₂ samples*

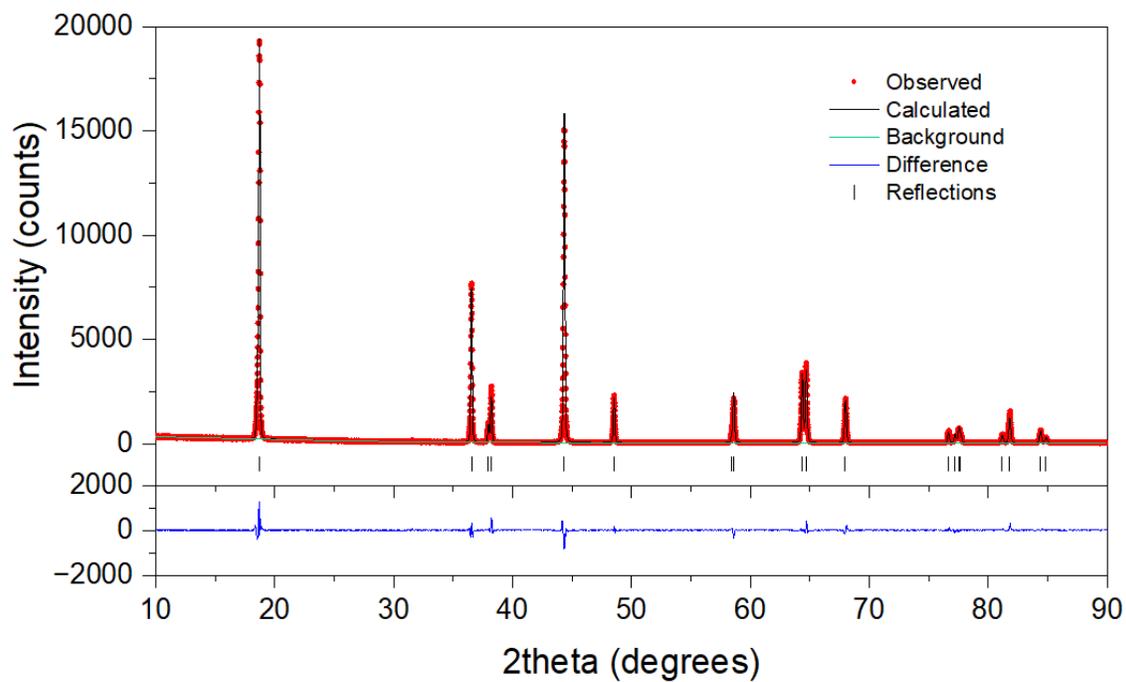

Figure S3. X-ray diffraction data for LiNiO$_2$.

Table S1. Rietveld refinement parameters for LiNiO$_2$.

|    | x     | y     | z        | Occupancy | U$_{iso}$ |
|----|-------|-------|----------|-----------|-----------|
| Li | 0.000 | 0.000 | 0.500    | 0.98(2)   | 0.006(2)  |
| Ni | 0.000 | 0.000 | 0.500    | 0.02(2)   | 0.013(2)  |
| Ni | 0.000 | 0.000 | 0.000    | 1.00      | 0.013(2)  |
| O  | 0.000 | 0.000 | 0.260(1) | 1.00      | 0.021(1)  |
| Space Group: R-3m   a = 2.879(1)   c = 14.204(1) ||||||
| G.O.F. = 1.38   R$_w$ = 9.13 ||||||

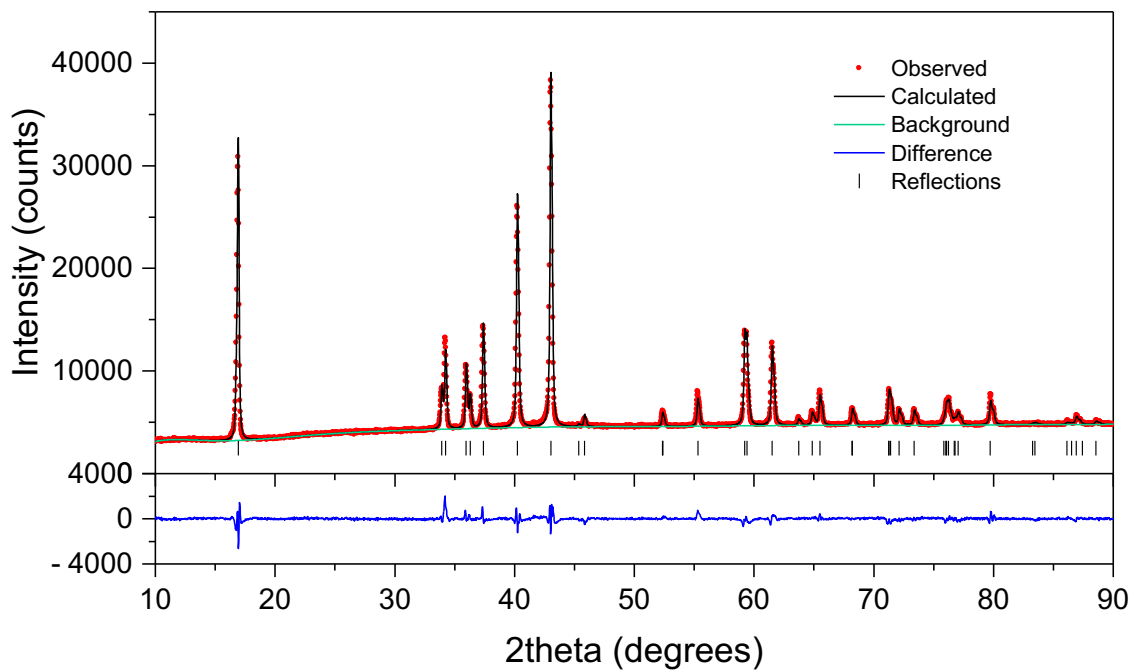

Figure S4. X-ray diffraction data for $NaNiO_2$.

Table S2. Rietveld refinement parameters for $NaNiO_2$.

|  | x | y | z | Occupancy | $U_{iso}$ |
|---|---|---|---|---|---|
| Na | 0.000 | 0.500 | 0.500 | 1.00 | 0.007(1) |
| Ni | 0.000 | 0.000 | 0.000 | 1.00 | 0.031(1) |
| O | 0.281(1) | 0.000 | 0.791(1) | 1.00 | 0.015(1) |
| Space Group: C2/m    a = 5.327(1)   b = 2.847(1)   c = 5.587(1) ||||||
| G.O.F. = 1.83    $R_w$ = 2.59 ||||||

## Calculated Resonant Inelastic X-ray Scattering Maps

RIXS maps calculated from the charge-transfer multiplet model above are shown in Figure S5. The scattered intensity near 857 eV photon energy and 2 eV loss is attributable to $S = 0$ Ni species.

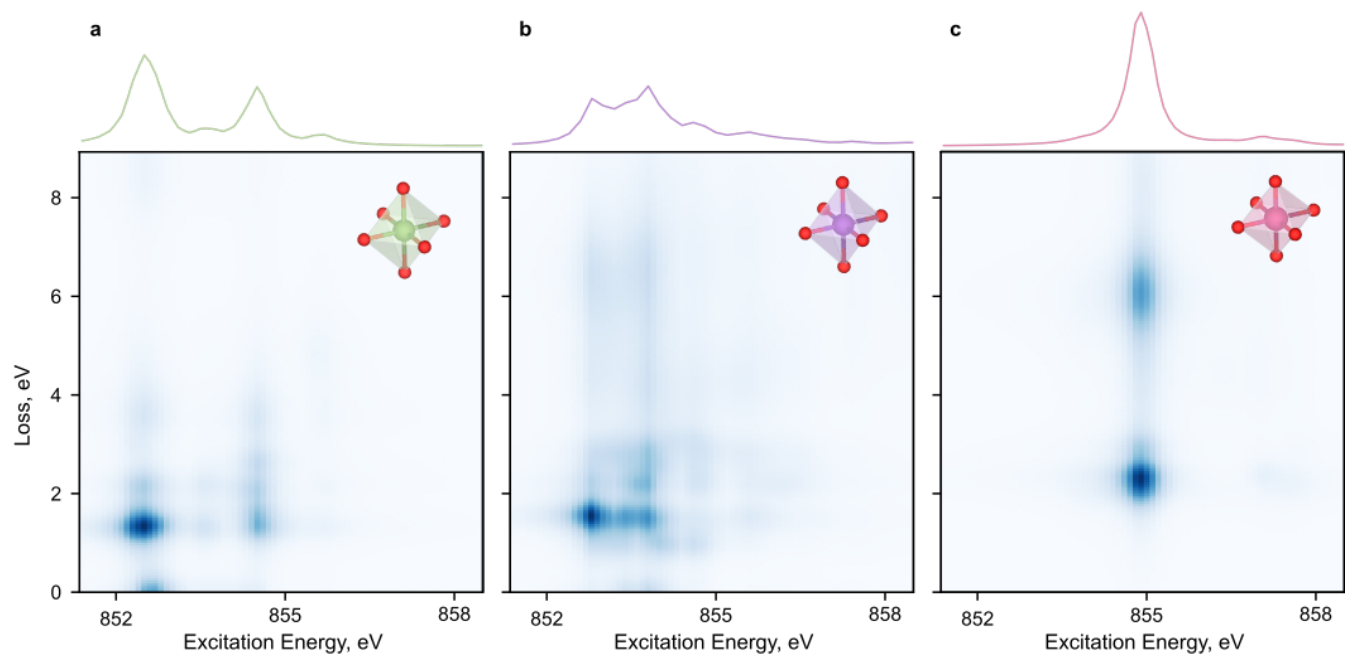

**Figure S5: Calculated localized transitions in the Ni $L_3$ edge RIXS of $LiNiO_2$.** (a) $S = 1$ Ni species, (b) $S = ½$ Ni species, (c) $S = 0$ Ni species. The integrated spectra are plotted above the intensity maps for each panel.

## Choice of density functional using thermodynamic data

The geometries of NiO and $LiNiO_2$ simulated in small cells are close to experimental values[11,70] and depend weakly on the choice of meta-GGA functionals (Table S3). The predicted thermodynamics and the geometry of $NiO_2$ depend more significantly on the choice of functional. Less expensive approaches such as PBE+U were discarded for their inability to accurately represent the $c$ lattice contraction of the delithiated $NiO_2$ phase[71]. Even at the meta-GGA level of theory, a dispersion correction is necessary to reproduce the $c$ lattice contraction of $NiO_2$. Since fitted Hubbard $U$ values are specific to oxidation states[72], empirical +$U$ corrections were avoided. In principle, the unusually large low-temperature heat capacity of $LiNiO_2$ makes a larger than normal contribution to room-temperature energies at 0.15 eV/f.u. for heating from 20 K to 300 K[73], but this correction was not applied here. SCAN over-stabilizes the NiO and $LiNiO_2$ phases, whereas $r^2$SCAN offers a precise match to NiO thermodynamics, but both are used to perform ab initio molecular dynamics shown below. Overall, $r^2$SCAN+rVV10 best reproduced the thermodynamics of $LiNiO_2$ and its constituent phases[74,75] and the geometry of the delithiated $NiO_2$ without major outlier errors.

**Table S3**: Comparison between structural and thermodynamic parameters of LiNiO$_2$ simulated with meta-GGA functionals and their experimental values.

| Parameter | SCAN | SCAN +rVV10 | r$^2$SCAN | r$^2$SCAN +rVV10 | Exp't | Ref. |
|---|---|---|---|---|---|---|
| $R\bar{3}m$ $a$, Å | 2.85 | 2.84 | 2.85 | 2.85 | 2.878 | 11 |
| $R\bar{3}m$ $c$, Å | 14.10 | 14.05 | 14.10 | 14.04 | 14.19 | 11 |
| $R\bar{3}m$ $\Delta H$, eV | -6.62 | -6.74 | -6.23 | -6.40 | -6.15 | 74 |
| NiO $a$, Å | 4.155 | 4.146 | 4.16 | 4.15 | 4.17 | 70 |
| NiO $\Delta H$, eV | -2.73 | -2.80 | -2.41 | -2.51 | -2.48 | 74 |
| Li$_2$O $\Delta H$, eV | -6.16 | -6.21 | -6.04 | -6.09 | -6.20 | 74 |
| NiO$_2$ $a$, Å | 2.77 | 2.77 | 2.78 | 2.78 | 2.81 | 76 |
| NiO$_2$ $c$, Å | 13.67 | 13.08 | 13.63 | 13.02 | 13.3 | 76 |

### Disproportionation in geometry relaxations

The relaxed energy of the commonly computed zigzag P2$_1$/c phase is 83 meV/f.u. below the bulk-average $R\bar{3}m$ phase, consistent with previous studies[19]. However, in relaxations starting with the $R\bar{3}m$ phase we found a partial disproportionation of Ni spins from $S$ = ½ to $S$ = 1 and $S$ = 0 for both SCAN and r$^2$SCAN meta-GGA functionals. If all $S$ = ½ Ni species are consumed and the resulting ($S$ = 1, $S$ = 0) pairs ordered linearly, such disproportionation yields the P2/c phase, previously predicted computationally[19,21,77]. The similarity between the three interpenetrating transition-metal sublattices in the ground state of Li(NiMnCo)O$_2$[32], the ground state of noncollinear spin models[33], and the disproportionated structure of AgNiO$_2$[13,14], which each exhibit three interpenetrating sublattices within the hexagonal layer of TMO$_6$ octahedra, suggests that the logical limit for disproportionation could be a phase where Ni species with spins $S$ = 1, $S$ = ½, and $S$ = 0 similarly occupy three sublattices, a static version of which was proposed by Foyevtsova et al.[21].

When this asymptotic three-fold disproportionated structure is relaxed, the volumes of the NiO$_6$ octahedra correlate with the spins of the Ni species: 10.9 Å$^3$ for $S$ = 1, 9.9 Å$^3$ for $S$ = ½, and 9.0 Å$^3$ for $S$ = 0 (Figure 2a), similar to AgNiO$_2$. All octahedra are somewhat distorted. The longest relaxed Ni–O bond is 2.06 Å, which is shorter than the Ni–O distance of 2.10 Å relaxed in the P2$_1$/c structure, and consistent with structural probes[29]. The relaxed limiting three-fold disproportionated structure and the two-fold disproportionated P2/c structure have energies 14 meV/f.u. and 12 meV/f.u. above the zigzag P2$_1$/c structure, respectively. The energetic penalty for changing the ordering (clockwise vs anti-clockwise) of the three sublattices within adjacent NiO$_2$ layers is 5 meV/f.u., and antiferromagnetic (AFM) or ferrimagnetic (FiM) orderings carry penalties of about 1 meV/f.u.

Configurational entropy is expected to favor the three-fold disproportionated structure and other intermediate compositions over the fully comproportionated (P2$_1$/c or NaNiO$_2$-like monoclinic C2/m) and two-fold disproportionated P2/c endpoints. The cluster expansion model used in Figure 2d identifies the comproportionated all-spin-half LiNiO$_2$ as 11 meV/f.u. lower in energy than the three-fold disproportionated structure but does not predict it to be observable. The full exploitation of the cluster expansion model is the subject of follow-on work.

## Dependence of Ni spin disproportionation on the choice of density functional

In the AIMD simulations of bulk pristine LiNiO$_2$, or bulk LiNiO$_2$ with an antisite Ni$_{Li}$ defect, substantial variation in spin was observed only for Ni. Spins on oxygens never exceeded ≈ 0.145 $\mu_B$. However, the temperature dependence of the free energy as projected onto the Ni spin coordinate differed substantially with the choice of k-points and meta-GGA functionals.

First, we compare the free-energy surfaces from the reference simulation performed with a 2×2×2 k-point mesh (Figure S6a, blue) versus a simulation performed at the Γ point exclusively (Figure S6a, pink) with the r$^2$SCAN functional. The two free-energy surfaces are drastically different: the Γ-point simulation does not predict stable disproportionation or any thermally activated dynamics. In the Γ-point simulation, magnetic moments below 0.2 $\mu_B$ or above 1.5 $\mu_B$ are not realized. Figure S6a also shows that a small distortion of the cell angles acquired during the relaxation of the three-spin disproportionated structure perturbs the energy surface (Figure S6a, purple), making transitions slightly rarer relative to the symmetric hexagonal high-temperature cell (Figure S6a, blue). Performing the simulation in the isobaric ensemble has a similar effect (Figure S6a, green), and using a larger cell with 144 atoms instead of 108 as in all the other runs (Figure S6a, orange) predicts slightly smaller activation energies. Overall, the inclusion of off-zone-center k-points is critical to reproducing experimental observables.

Using the SCAN functional (Figure S6b) yields a nearly temperature-independent set of free-energy surfaces, in contrast with r$^2$SCAN. Using SCAN favors all-spin-half arrangements upon cooling, while r$^2$SCAN favors disproportionation at cryogenic temperatures. The results from all functionals (r$^2$SCAN and SCAN+rVV10 not shown) agree best at high temperatures and disagree most towards cryogenic temperatures. This can be further seen in the projections of free energy surfaces onto the NiO$_6$ octahedral volumes in Figure S6c for r$^2$SCAN+rVV10 and Figure S6d for SCAN. This behavior of the simulated dynamics follows the trends in matching the thermodynamics of NiO and LiNiO$_2$ (Table S3): the choice of functional dominates the thermodynamic ensemble, simulation cell size, and non-local dispersion corrections.

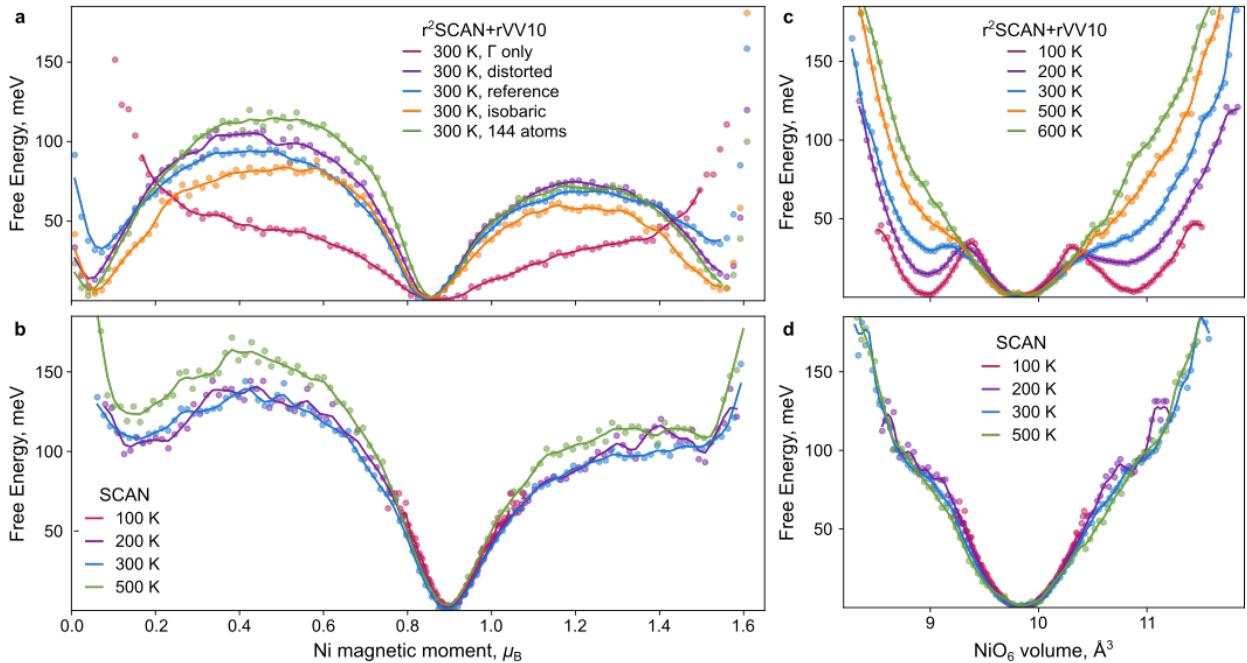

**Figure S6**: Sensitivity of simulated free energy surfaces to simulation parameters. (a) free energy surfaces versus the Ni magnetic moment with r$^2$SCAN+rVV10 at 300 K: reference as in Figure 3 (blue), relaxed cell with a small distortion (purple), hexagonal cell at the Γ point only (pink), hexagonal cell with 144 atoms (green), and simulation in the isobaric ensemble (orange). (b) free energy surfaces versus the Ni magnetic moment with SCAN and the hexagonal cell. (c-d) free energy surfaces versus the NiO$_6$ volume with r$^2$SCAN+rVV10 (c) or SCAN (d) and the hexagonal cell.

We hypothesize that the sensitivity to the choice of functional is due to the small relative energy differences under investigation (<20 meV/f.u. between all relevant relaxed structures) being dominated by phonon energies (>60 meV) and the differences in predicted thermodynamic formation energies across functionals (e.g., 0.3–0.4 eV/f.u. for NiO, Table S3). This may represent a limit of precision for computational studies of LiNiO$_2$ structure and dynamics.

### Structural Descriptors

The key observable structural signatures of local geometry are the unit cell parameters (Table S3) and the Ni-O bond length, which are typically probed by EXAFS[7,27,28] and diffraction[29]. These methods probe the average structure, and in both cases a fitting or refinement is required to obtain structural information. Several aspects complicate the use of structural information for fingerprinting Ni species: the primacy of the electronic coordinate, the noise in the structural coordinate, and the necessary regularization in interpreting experimental data. The electronic coordinate provides a stronger and lower-noise separation of the three Ni species (main text and Figure S6), and the NiO$_6$ octahedra of multiple species have the same bond lengths in static calculations. Despite drastically different predictions of Ni electronic speciation (Figure S6ab), the SCAN and r$^2$SCAN functionals yield similar unit cell geometries (Table S3) and distributions of Ni-O bond lengths (Figure S7). The low-temperature bond length distributions are sharpest and provide the strongest contrast between functionals. The r$^2$SCAN structures appear stiffer with narrower bond length distributions and have shorter long bonds. However, between 200 K and 300 K the two distinct bond populations merge so that by 300 K, the distributions of bond lengths are virtually indistinguishable if the simulations are carried out in cell geometries relaxed from the hexagonal unit cell (Figure S7c), despite differences in the predicted electronic structure. Both hexagonal simulations yield long bonds at ≈2.05 Å, in close agreement with experiment, for both electronic structures. By contrast, the zigzag structure ($P2_1/c$ in Figure S7c) predicts long bonds at ≈2.10Å, which is longer than any experimentally obtained ones.

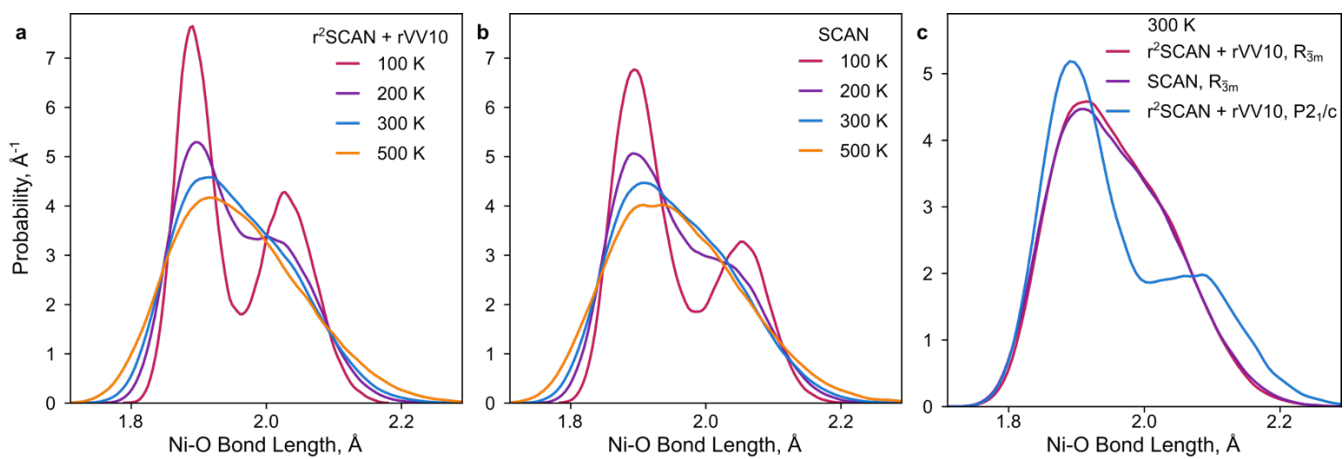

**Figure S7: Simulated Ni-O bond distances in LiNiO$_2$.** (a) simulated with r2SCAN+rVV10 as a function of temperature, (b) simulated with SCAN as a function of temperature (c) at 300 K for two unit cells.

## Spin Transition Statistics

To verify whether the AIMD trajectories were of sufficient length to construct free energy surfaces, transitions between spin states were counted for each Ni ion by milestoning at 200 K and above (Methods). Simulations were continued until every ion had exhibited multiple transitions (Figure S8). Excluding the initial 1 ps for thermostat equilibration, the simulations at 200 K and 300 K were collected for over 10 ps with the longest residence times 7.7 ps at 200 K and 2.7 ps at 300 K. The trajectories at 500 K and 600 K were collected for at least 4.6 ps with the longest residence times 1.4 ps. At 300 K, the fastest transition times are on the order of 50 fs, close to one period of the Raman-active vibrations of Ni–O bonds[78]. This is consistent with the coupling between $NiO_6$ geometry and Ni states: geometry changes mediated by vibrations set a speed limit on changes in electronic states.

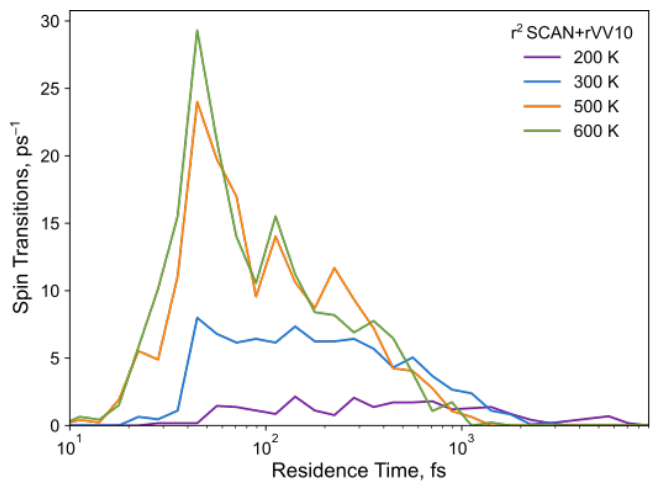

**Figure S8**: Spin transitions in ab initio molecular dynamics. The frequency of spin transitions is plotted versus the length of time the Ni ions spend in a particular spin state before making a transition.

## Defect chemistry and stabilization of the $Ni_{Li}$ antisite

We next explore the influence of disproportionation on defect chemistry via the localization of electronic charge carriers and the intrinsic antisite defect. Hole and electron polarons are relaxed in the limiting three-fold disproportionated structure by distorting the $S = ½$ Ni octahedra to sizes commensurate with $S = 0$ and $S = 1$, respectively. While distortions of neighboring octahedra are minimal, some relaxations instead yield a reordering of the Ni spins and their respective octahedral volumes. Results are consistent between 108-atom and 144-atom supercells. The stabilization of polarons is consistent with a correspondence between the spin states and formal charge states of Ni species. By contrast, we have not been able to stabilize a hole polaron or a Ni with $S ≈ 0$ in the $P2_1/c$ structure at the meta-GGA level of theory, except in a 2×3×2 supercell (96 atoms) and surrounded by six opposite-sign $S = -½$ spins on adjacent Ni ions, which forms a partial superlattice like the three-fold disproportionation discussed here. This arrangement is clearly artificial, is energetically unfavorable relative to a delocalized hole, and is not stabilized in other supercells. Electron polarons are stable as $S = 1$ Ni in the $P2_1/c$ structure, consistent with earlier work[62].

We now examine the $Ni_{Li}$ antisite defect. To calculate the absolute defect formation energies, we first determine the chemical potentials of the elements from reference phase energies[64] (Figure S9ab). Accounting for the paramagnetic transition of NiO and representing the energetics of NiO accurately (Table S3) are key to estimating $\mu_{Li}$ and $\mu_{Ni}$ and the defect formation energy. The phase equilibrium at synthesis is between LiNiO$_2$ and a disordered rocksalt, computationally predicted as LiNi$_2$O$_3$[79]. This is consistent with the experimental synthesis pathway from NiO via lithium insertion followed by the formation of ordered layers[80]. Finally, since layered Li$_2$NiO$_3$ is not easily synthesized[81], and not observed as a decomposition product, we include an additional region of likely metastability over Li$_2$NiO$_3$ (light grey in Figure S9ab).

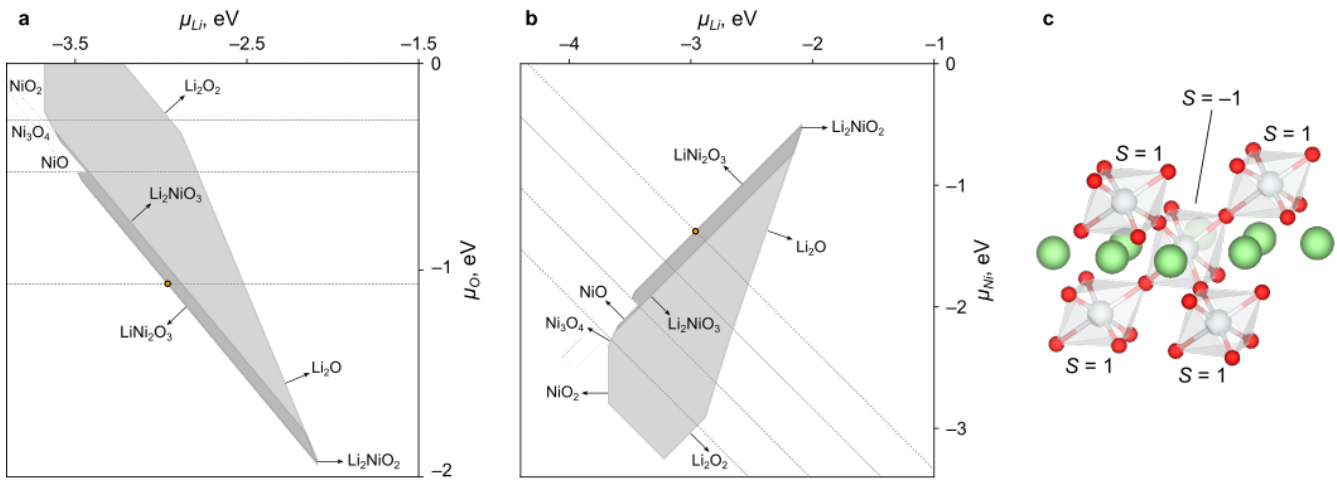

**Figure S9: Phase and defect stability in LiNiO$_2$.** (a,b) thermodynamic phase stability of disproportionated LiNiO$_2$, with competing phases highlighted, and the synthetic conditions denoted by an orange circle. (c) Antiferromagnetic stabilization of the Ni$_{Li}$ antisite defect by $S$ = 1 Ni ions.

Within disproportionated structures, the energy of the Ni$_{Li}$ antisite depends on the arrangements of the nickel spins adjacent to it within the NiO$_2$ layers. The lowest-energy configuration for the Ni$_{Li}$ defect is one where $S$ = –1 Ni$_{Li}$ is surrounded by $S$ = 1 Ni ions via 180-degree Ni–O–Ni motifs (Figure S9c), reminiscent of antiferromagnetic NiO. At synthesis conditions, the formation energies of multiple configurations are negative, with the lowest at –60 meV, whereas the lowest defect energy in the P2$_1$/c structure is 220 meV. In the P2$_1$/c structure, only one Ni$_{Ni}$ carries $S$ = 1 to compensate the introduction of the Ni$_{Li}$. Our estimate of the formation energies for the most stable local configurations of Ni$_{Li}$ is lower than the already low figures reported earlier[62] due to accounting for this additional antiferromagnetic stabilization enabled by disproportionation in LiNiO$_2$.

While the local atomic and spin arrangements are expected to fluctuate rapidly at synthesis, the results remain relevant for room-temperature degradation pathways during battery operation. In AIMD trajectories incorporating a Ni$_{Li}$ antisite, the Ni ions sharing octahedral corners with the defect as in Figure S9c carry $S$ = 1 on average 50% more often than Ni ions away from the defect: the $S$ = 1 that compensate the antisite are not free even at elevated temperatures. We conclude that magnetic stabilization contributes to the ubiquity of this pervasive defect. Ni$_{Li}$ is stable not only because of the size match between Ni and Li, but also because of the ability of the Ni to disproportionate in the layered structure, which distinguishes it from Co and Mn oxides, and from the monoclinic NaNiO$_2$. This effect likely contributes to the driving forces for surface reconstructions and Ni migration into the Li layer during charging.